# Phase separation in complex mixtures with many components: analytical expressions for spinodal manifolds.

Arjen Bot,[a,b] Erik van der Linden[b] and Paul Venema[b,*]

[a] *Unilever Foods Innovation Centre, Bronland 14, NL-6708 WH Wageningen, The Netherlands*

[b] *Laboratory of Physics and Physical Chemistry of Foods, Department of Agrotechnology and Food Sciences, Wageningen University and Research, Bornse Weilanden 9, NL-6708 WG Wageningen, The Netherlands*

*Corresponding author

## Abstract

The phase behavior is investigated for systems composed of a large number of macromolecular components $N$, with $N \geq 2$. Liquid-liquid phase separation is modelled using a virial expansion up to the second order of the concentrations of the components. Formal analytical expressions for the spinodal manifolds in $N$ dimensions are derived that simplify their calculation (by transforming the original problem into inequalities that can be evaluated numerically using linear programming techniques). In addition, a new expression is obtained to calculate the critical manifold and the composition of the co-existing phases. The present analytical procedure complements previous attempts to handle spinodal decomposition for many components using a statistical approach based on Random Matrix Theory. The results are relevant for predicting effects of polydispersity on phase behavior in fields like polymer or food science, and to liquid-liquid phase separation in the cytosol of living cells.

## Keywords:

liquid-liquid phase separation (LLPS), segregation, association, condensation, virial expansion, second virial coefficient, polymers, co-existing phases, binodal, spinodal, critical point, Lambert-W function, cytosol

## Introduction

Phase separation in mixtures of molecules is a phenomenon leading to the formation of distinct domains on a meso- and/or macroscopic scale, which differ in molecular composition. This phenomenon is described in terms of equilibrium thermodynamics of the system and may occur at sufficiently high concentrations of the components. Phase separation plays a role in classical fields as diverse as polymer and food technology, hematology, wastewater treatment, archaeology, and forensics. Considerable effort was invested in experimental and computational work on multicomponent mixtures [1,2,3,4,5,6,7,8,9]. In addition to the abovementioned areas of interest, phase separation in living soft matter is currently a highly active field of research, in particular in relation to the cytosol of cells containing complex mixtures of thousands of components [10,11,12,13,14].

When describing such mixtures, one might ignore the complexity of the problem by assuming that complex mixtures can be approximated by a mixture of a small number of monodisperse components, each with distinct physical properties. This approach is often taken for binary or ternary mixtures of polydisperse polymers [15].

Alternatively, one may embrace the full complexity of the system, and treat every component as unique (either in terms of its chemical or physical properties). This last route requires an approach that addresses mixtures of many components and involves solving a large set of non-linear algebraic equations simultaneously. However, the numerical evaluation of such sets of equations is far from trivial, which has resulted in the introduction of alternative methodologies. The first of such a methodology was presented two decades ago by Sear and Cuesta [16] and was based on Random Matrix Theory (RMT) [17]. This seminal work, valid for mixtures of a very large number of components, prompted various alternative approaches for describing phase behavior of complex mixtures, culminating in a field that was reviewed recently by Jacobs [18] and Pappu et al. [19]. As will be shown in the Results section, RMT requires that the number of components fulfills $N \gtrsim 10^3$. This identifies a gap in the numerical approaches for a low number of components and RMT, where the latter is only valid for a very high number of components.

An approach for moderately large numbers of components (typically $2 \leq N \lesssim 10^3$) is to generalise previously obtained (mostly) analytical results for the key characteristics of the phase diagrams for binary [20,21,22,23,24] and ternary mixtures [25] towards mixtures containing $N$ components. Although this seems challenging, surprisingly, this is found to be viable route, as is discussed in the remainder of this paper. The present analysis for $N$ component mixtures is simplified by introducing three sets of parameters. One set of parameters can be related to the tangents to the critical manifold, another set to the slopes of the tie-lines connecting the co-existing phases, and the third set can be considered as free parameters. The analysis gives formal analytical expressions for the spinodal manifold without being restricted to $N \gtrsim 10^3$, thus closing the aforementioned gap between the existing methods for low and very high numbers of components. Two additional advantages of the approach in this paper are that it demonstrates a route to determine the critical manifolds, and the compositions of co-existing phases for any number $N$ of components.

The present paper focusses mainly on developing the methods involved in establishing the phase behavior for mixtures of many components and less on performing extensive numerical calculations.

# Methods: theory

## Co-existence equations for $N$ components and $P$ phases.

For $N$ components and two phases ($P$=2) the expression for the Helmholtz free energy $F$ ($J$) can be approximated by a virial expansion in terms of molar concentrations including terms up to the second order in molar concentration as

$$\frac{F}{RTV} = \sum_{i=1}^{N} c_i \cdot ln\, c_i + \sum_{i,j=1}^{N} B_{ij} c_i c_j \tag{1}$$

where $B_{ij} \equiv B_{ji}$. The expressions for the osmotic pressure and chemical potentials can be derived as

$$\frac{\Pi}{RT} = -\frac{1}{RT}\left(\frac{\partial F}{\partial V}\right)_{T,n_i} = \sum_{i=1}^{N} c_i + \sum_{i,j=1}^{N} B_{ij} c_i c_j \tag{2}$$

$$\frac{\mu_i}{RT} = \frac{1}{RT}\left(\frac{\partial F}{\partial n_i}\right)_{T,V,n_{j\neq i}} = ln\, c_i + 2\sum_{j=1}^{N} B_{ij} c_j + 1 \qquad \text{with } i = 1,2,\dots,N \tag{3}$$

where $n_i$ $(mol)$ is the number of moles and $c_i = n_i/\mathrm{V}$ the molar concentration of component $i$, $T$ $(K)$ the absolute temperature, $R$ $(J \cdot K^{-1} \cdot mol^{-1})$ the gas constant, $V$ $(m^3)$ the total volume of the system, $B_{ii}$ $(m^3 \cdot mol^{-1})$ the second virial coefficient of polymer $i$ ($i$=1,2,….,$N$) and $B_{ij}$ the second cross virial coefficient for polymers $i$ and $j$ ($i,j$=1,2,….,$N$; $i{\neq}j$) that can be any real number. These equations form the basis for the thermodynamic description of phase behavior. In the present analysis each of the $N\geq2$ components is regarded as a macromolecular component, where the solvent is integrated out. Alternatively, it is possible to add the solvent explicitly as an additional component (cf. Ref 26), provided that the interactions are correctly taken into account via the second virial coefficients. Without loss of generality, the discussion can be limited to the case of two separated phases in thermodynamic equilibrium, $I$ and $II$, where the osmotic pressure and all $N$ chemical potentials need to be the same in each phase

$$\frac{\Pi^{I}}{RT} = \frac{\Pi^{II}}{RT} \tag{4}$$

$$\frac{\mu_i{}^I}{RT} = \frac{\mu_i{}^{II}}{RT} \qquad \text{with } i = 1,2,\dots,N \tag{5}$$

Using Eqs. (2)-(3), the explicit expressions are given by

$$\sum_{i=1}^{N} c_i^I + \sum_{i,j=1}^{N} B_{ij} c_i^I c_j^I = \sum_{i=1}^{N} c_i^{II} + \sum_{i,j=1}^{N} B_{ij} c_i^{II} c_j^{II} \tag{6}$$

$$ln\, c_i^I + 2\sum_{j=1}^{N} B_{ij} c_j^I + 1 = ln\, c_i^{II} + 2\sum_{j=1}^{N} B_{ij} c_j^{II} + 1 \qquad \text{with } i = 1,2,\dots,N \tag{7}$$

For mixtures where *P* >2, the equalities in Eqs (4) and (5) should be extended to include more phases, identified by *III*, *IV*, etc. The maximum value for *P* is determined by the Gibbs phase rule. Following an earlier approach [23] it is convenient to introduce the parameter $S_{m,ij}$, the slope of a tie-line in the (*i*,*j*) plane multiplied by -1, defined by

$$S_{m,ij} \equiv -\frac{c_i^{II} - c_i^I}{c_j^{II} - c_j^I} = -\frac{1}{S_{m,jk} S_{m,ki}} \equiv \frac{1}{S_{m,ji}} \qquad \text{with } i,j = 1,2,\dots,N \tag{8}$$

where $(c_1^I, c_2^I, \dots, c_N^I)$ and $(c_1^{II}, c_2^{II}, \dots, c_N^{II})$ are the compositions of the two co-existing phases. $S_{m,ij}$ may vary in the range <-∞,∞>. The third term in Eq (8) can be confirmed by substituting the definition for the parameters $S_{m,jk}$ and $S_{m,ki}$. Note that in case of segregative phase separation, each phase is enriched in one of the components and therefore $S_{m,ij} > 0$. Oppositely for associative phase separation, where one of the phases is enriched and the other phase depleted in both components, and therefore $S_{m,ij} < 0$. In addition, there are *N* parameters $S_{m,ij}$ in Eq (8), of which (*N*-1) parameters can be chosen freely because $S_{m,ii} = -1$ is fixed. Eqs. (6)-(7) consist of (*N*+1) equations with *2N* unknowns. This leads to a set of 2*N* equations with 2*N* unknowns. Substituting Eq (8) in Eqs. (6)-(7) results in

$$\sum_{i=1}^{N} S_{m,i1} \left( \left( \sum_{j=1}^{N} B_{ij} S_{m,ji} \right) (c_i^I + c_i^{II}) - 1 \right) = 0 \tag{9}$$

$$ln\left( \left( 2\sum_{j=1}^{N} B_{ij} S_{m,ji} \right) c_i^I \right) - \left( 2\sum_{j=1}^{N} B_{ij} S_{m,ji} \right) c_i^I = ln\left( \left( 2\sum_{j=1}^{N} B_{ij} S_{m,ji} \right) c_i^{II} \right) - \left( 2\sum_{j=1}^{N} B_{ij} S_{m,ji} \right) c_i^{II} \qquad \text{with } i = 1,2,\dots,N \tag{10}$$

Defining the constants $c_{i,s}$ as

$$c_{i,s} = \frac{1}{2\left(\sum_{j=1}^{N} B_{ij} S_{m,ji}\right)} \qquad \text{with } i = 1,2,\dots,N \tag{11}$$

allows writing Eq (9) as

$$\sum_{i=1}^{N} S_{m,i1}\left(\left(\frac{c_i^I}{c_{i,s}}-1\right)+\left(\frac{c_i^{II}}{c_{i,s}}-1\right)\right)=0 \tag{12}$$

and Eq (10) as

$$\ln\left(\frac{c_i^I}{c_{i,s}}\right)-\frac{c_i^I}{c_{i,s}}=\ln\left(\frac{c_i^{II}}{c_{i,s}}\right)-\frac{c_i^{II}}{c_{i,s}} \qquad \text{with } i=1,2,\dots,N \tag{13}$$

with -by definition- the solutions [21,23,25]

$$W\left(-(c_i^I/c_{i,s})e^{-(c_i^I/c_{i,s})}\right)=W\left(-(c_i^{II}/c_{i,s})e^{-(c_i^{II}/c_{i,s})}\right) \qquad \text{with } i=1,2,\dots,N \tag{14}$$

where *W* refers to the Lambert-*W* function [27]. The expressions in Eq (11) normalise the coordinates in Eqs (12) and (13). Eq (14) links the concentration of each component *i* in phase *I* to that in phase *II*. For real arguments, the Lambert-*W* function has zero, one or two solutions, corresponding to isotropic mixing, the location of the critical points, and (segregative and/or associative) phase separation, respectively. This makes the Lambert-*W* function eminently suited to describe solutions to phase separation problems. In case of two solutions, the solution for each component *i* in phase *I* and *II* is located either on the $W_{-1}$-branch or $W_0$-branch of the Lambert-*W* function. The $W_{-1}$-branch refers to high concentrations and the $W_0$-branch to low concentrations. In case of segregative phase separation (where one phase is enriched in component *i* and the other phase enriched in component *j* and $S_{m,ij} > 0$) the co-existing phases for components *i* and *j* correspond to $\left(c_i^I, c_j^I\right)=(W_{-1}, W_0)$ and $\left(c_i^{II}, c_j^{II}\right)=(W_0, W_{-1})$, where the arguments of the Lambert-*W* functions (different for each coordinate *i* or *j*) were omitted for clarity. In case of associative phase separation or condensation (where one phase is enriched and the other phase depleted in both components *i* and *j*, and $S_{m,ij} < 0$) the co-existing phases correspond to $\left(c_i^I, c_j^I\right)=(W_0, W_0)$ and $\left(c_i^{II}, c_j^{II}\right)=(W_{-1}, W_{-1})$. The $S_{m,ij}$ can be any real number, but note that the physically relevant non-negativity of coordinates $c_{i,s}$ for a specific choice of $B_{ij}$ and $S_{m,ij}$ in Eq (11) needs to be confirmed separately. Eq (8) can be written as:

$$\left(\frac{c_i^I}{c_{i,s}}-1\right)-\left(\frac{c_i^{II}}{c_{i,s}}-1\right) = -S_{m,ij}\frac{c_{j,s}}{c_{i,s}}\left(\left(\frac{c_j^I}{c_{j,s}}-1\right)-\left(\frac{c_j^{II}}{c_{j,s}}-1\right)\right) \qquad \text{with } i,j=1,2,\dots,N \tag{15}$$

and can be combined with Eq (12) to create a matrix equation (see Appendix A)

$$\begin{pmatrix} \frac{c_1^I}{c_{1,s}}-1 \\ \frac{c_2^I}{c_{2,s}}-1 \\ \vdots \\ \frac{c_N^I}{c_{N,s}}-1 \end{pmatrix} = \boldsymbol{V}\begin{pmatrix} \frac{c_1^{II}}{c_{1,s}}-1 \\ \frac{c_2^{II}}{c_{2,s}}-1 \\ \vdots \\ \frac{c_N^{II}}{c_{N,s}}-1 \end{pmatrix} \tag{16}$$

with the elements of the $i$-th row and $j$-th column of the $NxN$ matrix $\boldsymbol{V}$ defined as

$$V(i,j) = \frac{\left(\sum_{k=1}^{N}\left(\frac{S_{m,k1}^2}{c_{k,s}}\right)\right)\delta_{ij} - 2\frac{S_{m,i1}S_{m,j1}}{c_{i,s}}}{\sum_{k=1}^{N}\left(\frac{S_{m,k1}^2}{c_{k,s}}\right)} \tag{17}$$

$$= \delta_{ij} - \frac{2}{\sum_{k=1}^{N}\left(\frac{S_{m,ki}S_{m,kj}}{c_{k,s}/c_{i,s}}\right)} \qquad \text{with } i,j = 1,2,\dots,N$$

with $\delta_{ij}$ the Kronecker delta ($\delta_{ij} = 1$ for $i$=$j$ and $\delta_{ij} = 0$ for $i$≠$j$). The binodal manifold can be calculated from Eqs (16)-(17) together with Eq (14) (cf. Refs 23 and 25), and has dimension ($N$-1) (in line with the number of $S_{m,ij}$ that can be chosen freely). The expressions have been simplified from 2$N$ concentrations $c_i^I$ and $c_i^{II}$ (Eqs (6)-(7)) to $N$ concentrations $c_i^I$ plus ($N$-1) parameters $S_{m,ij}$ (because $S_{m,ii} = -1$), effectively reducing the number of equations from to 2$N$ to 2$N$-1. For $N$=3, Eqs (16)-(17) reduce to Eq (52) in Ref 25. For $N$=2, Eq (48) in Ref 23 was written for different coordinate vectors, but this expression can be shown to be equivalent to Eq (16) and (17).

The composition of $P$ co-existing phases for $N$ components can be obtained from the Lambert-$W$ functions (cf Eq (14)).

$$W\left(-\frac{c_i^I}{c_{i,s}}e^{-\frac{c_i^I}{c_{i,s}}}\right) = W\left(-\frac{c_i^{II}}{c_{i,s}}e^{-\frac{c_i^{II}}{c_{i,s}}}\right) = \dots = W\left(-\frac{c_i^P}{c_{i,s}}e^{-\frac{c_i^P}{c_{i,s}}}\right) \qquad \text{with } i = 1,2,\dots,N \tag{18}$$

As noted earlier, this allows for several solutions per component, potentially located either on the same branch or on two different branches of the Lambert-$W$ function, where the physically relevant solution fulfils Eqs (4) and (5) and all concentrations are positive.

## Spinodal manifolds for $N$ components.

The local curvature of the Helmholtz free energy $F$ manifold is given by the Hessian matrix $\boldsymbol{M_1}$, and characterises the local stability of a mixture against phase separation. Element $\boldsymbol{M_1}(i,j) = \frac{1}{RT}\left(\frac{\partial\mu_i}{\partial c_j}\right)_{T,V,c_{k\neq j}}$ is given by

$$\boldsymbol{M_1}(i,j) = \frac{\delta_{ij}}{c_i} + 2B_{ij} \qquad \text{with } i,j = 1,2,\dots,N \tag{19}$$

with $\delta_{ij}$ the Kronecker delta and $B_{ij} \equiv B_{ji}$, making $\boldsymbol{M_1}$ symmetrical. Based on previous results for $N$=2 and $N$=3 [23,25] , it is possible to find an analytical expression for the coordinates $c_{i,sp}$ of the spinodal manifold for $N$ components (the index '$sp$' refers to spinodal):

$$c_{i,sp} = \frac{1}{2\left(\sum_{j=1}^{N} B_{ij} S_{sp,ji}\right)} \quad \text{with } i = 1,2,\dots,N \tag{20}$$

where $S_{sp,ji}$ is a parameter satisfying the relation

$$S_{sp,ji} = -\frac{1}{S_{sp,ik} S_{sp,kj}} = \frac{1}{S_{sp,ij}} \qquad \text{with } i,j = 1,2,\dots,N \tag{21}$$

Eq (21) has the same form as Eq (8). It is shown in Appendix B that Eq (20) indeed satisfies the requirement for the spinodal [28], i.e. $Det\ \boldsymbol{M_1} = 0$ (where *Det* refers to the determinant), when Eqs (20) and (21) are substituted in matrix $\boldsymbol{M_1}$ (Eq (19)). It follows that Eq (20) represents the coordinates for the (*N*-1) dimensional spinodal manifold, characterised by the (*N*-1) free parameters $S_{sp,ji}$ (with $i \neq j$) and one fixed parameter $S_{sp,ii} \equiv -1$. Note that the fixed value of parameter $S_{sp,ii}$ warrants that the spinodal manifold has a dimension of (*N*-1), preventing that every point in composition space could be written as a point on the spinodal. The non-negativity of coordinates $c_{i,sp}$ for a specific choice of $B_{ij}$ and $S_{sp,ji}$ in Eq (20) needs to be checked separately, as the existence of such unphysical solutions is inherent to the mathematical model (cf. Fig. 1 in Ref 21 for binary mixtures). The procedure to check for non-negativity will be illustrated in the Results section. It is noted that spinodal manifolds obtained via $Det\ \boldsymbol{M_1} = 0$ do not always separate unstable from (meta-) stable regions [29] (cf. Fig 2f in Ref 25). For *N*=3, Eq (20) reduces to the previously obtained spinodal manifold for ternary mixtures [25] (See also Table 1). For *N*=2, the parameter $S_{sp}$ in Ref 23 is related to $S_{sp,21}$ in the present paper by $\sqrt{S_{sp}} = (B_{12} S_{sp,21} - B_{11})/(B_{12} - B_{22} S_{sp,21})$ on the spinodal and therefore to $S_{sp} = S_{sp,21}$ in the critical point.

## Critical manifolds for *N* components.

The standard procedure to evaluate the critical manifold involves solving $Det\ \boldsymbol{M_1} = 0 \wedge Det\ \boldsymbol{M_2} = 0$ [28], where matrix $\boldsymbol{M_2}$ is constructed by replacing the bottom row (or any other row for this matter) of matrix $\boldsymbol{M_1}$ by [28,30,31]

$$\left[\frac{\partial}{\partial c_1}(Det\ \boldsymbol{M_1}) \quad \frac{\partial}{\partial c_2}(Det\ \boldsymbol{M_1}) \quad \dots \quad \frac{\partial}{\partial c_N}(Det\ \boldsymbol{M_1})\right] \tag{22}$$

where $\frac{\partial}{\partial c_i}$ is the partial derivative relative to $c_i$ (and the resulting matrix is shown in Eq (25)). The following relation holds

$$\frac{\partial}{\partial c_j}(Det\ \boldsymbol{M_1}) = -\left\{Det\ \boldsymbol{M}_1^{(j,j)}\right\}\frac{1}{{c_{j,c}}^2} = -{S_{c,1j}}^3\left\{Det\ \boldsymbol{M}_1^{(j,j)}\right\}\frac{{S_{c,j1}}^3}{{c_{j,c}}^2} \qquad \text{with } j = 1,2,\dots,N \tag{23}$$

with $\boldsymbol{M}_1^{(j,j)}$ the minor of $\boldsymbol{M_1}$ relative to the element (*j*,*j*) and where use was made of the following relation for $S_{c,ji}$, the tangent to the spinodal manifold at the location of the binodal manifold in the (*i*,*j*) plane multiplied by -1,

| | ***N*=2** <br> *i*=1,2 | ***N*=3** <br> *i*=1,2,3 | ***N*** <br> *i*=1,2,...,*N* |
|---|---|---|---|
| | Refs 21,23,24 | Ref 25 | Present work |
| Spinodal points | $c_{i,sp} = \frac{1}{2\left(B_{i1}S_{sp,1i} + B_{i2}S_{sp,2i}\right)}$ | $c_{i,sp} = \frac{1}{2\left(B_{i1}S_{sp,1i} + B_{i2}S_{sp,2i} + B_{i3}S_{sp,3i}\right)}$ | $c_{i,sp} = \frac{1}{2\left(\sum_{j=1}^{N} B_{ij}S_{sp,ji}\right)}$ |
| Critical point(s) | $c_{i,c} = \frac{1}{2\left(B_{i1}S_{c,1i} + B_{i2}S_{c,2i}\right)}$ <br> $S_{c,21} = \left(\frac{c_{2,c}}{c_{1,c}}\right)^{2/3}$ | $c_{i,c} = \frac{1}{2\left(B_{i1}S_{c,1i} + B_{i2}S_{c,2i} + B_{i3}S_{c,3i}\right)}$ <br> $\frac{S_{c,11}^3}{c_{1,c}^2} + \frac{S_{c,21}^3}{c_{2,c}^2} + \frac{S_{c,31}^3}{c_{3,c}^2} = 0$ | $c_{i,c} = \frac{1}{2\left(\sum_{j=1}^{N} B_{ij}S_{c,ji}\right)}$ <br> $\sum_{j=1}^{N} \frac{S_{c,j1}^3}{c_{j,c}^2} = 0$ |
| Co-existence equations | $c_{i,s} = \frac{1}{2\left(B_{i1}S_{m,1i} + B_{i2}S_{m,2i}\right)}$ <br> $\sum_{i=1}^{2} S_{m,i1}\left(\left(\frac{c_i^I}{c_{i,s}} - 1\right) + \left(\frac{c_i^{II}}{c_{i,s}} - 1\right)\right) = 0$ <br> $W\left(-\frac{c_i^I}{c_{i,s}} e^{-\frac{c_i^I}{c_{i,s}}}\right) = W\left(-\frac{c_i^{II}}{c_{i,s}} e^{-\frac{c_i^{II}}{c_{i,s}}}\right)$ | $c_{i,s} = \frac{1}{2\left(B_{i1}S_{m,1i} + B_{i2}S_{m,2i} + B_{i3}S_{m,3i}\right)}$ <br> $\sum_{i=1}^{3} S_{m,i1}\left(\left(\frac{c_i^I}{c_{i,s}} - 1\right) + \left(\frac{c_i^{II}}{c_{i,s}} - 1\right)\right) = 0$ <br> $W\left(-\frac{c_i^I}{c_{i,s}} e^{-\frac{c_i^I}{c_{i,s}}}\right) = W\left(-\frac{c_i^{II}}{c_{i,s}} e^{-\frac{c_i^{II}}{c_{i,s}}}\right)$ | $c_{i,s} = \frac{1}{2\left(\sum_{j=1}^{N} B_{ij}S_{m,ji}\right)}$ <br> $\sum_{i=1}^{N} S_{m,i1}\left(\left(\frac{c_i^I}{c_{i,s}} - 1\right) + \left(\frac{c_i^{II}}{c_{i,s}} - 1\right)\right) = 0$ <br> $W\left(-\frac{c_i^I}{c_{i,s}} e^{-\frac{c_i^I}{c_{i,s}}}\right) = W\left(-\frac{c_i^{II}}{c_{i,s}} e^{-\frac{c_i^{II}}{c_{i,s}}}\right)$ |
| Common rules for parameters | $S_{m,ij} \equiv -\frac{c_i^{II} - c_i^I}{c_j^{II} - c_j^I}$ <br> $S_{sp,ii} = S_{c,ii} = S_{m,ii} = -1$ <br> $S_{sp,ij} = -\frac{1}{S_{sp,jk}S_{sp,ki}} = \frac{1}{S_{sp,ji}}$; $S_{c,ij} = -\frac{1}{S_{c,jk}S_{c,ki}} = \frac{1}{S_{c,ji}}$; $S_{m,ij} = -\frac{1}{S_{m,jk}S_{m,ki}} = \frac{1}{S_{m,ji}}$; | | |

Table 1: Comparison between the model results obtained for *N*=2, 3 and *N*. Note that $B_{ij} \equiv B_{ji}$. The Table above describes the case for two separate phases *I* and *II* (*P*=2), but can be extended easily to include more phases by adding expressions for phase *III*, *IV*, etc. The function *W* refers to the Lambert-*W* function [27], which can be calculated straightforwardly using modern software packages.

$$S_{c,ij} = -\frac{1}{S_{c,jk}S_{c,ki}} = \frac{1}{S_{c,ji}} \qquad \text{with } i,j = 1,2,\ldots,N \tag{24}$$

with $S_{c,ii} \equiv -1$. Eq (24) corresponds to the limit of Eq (8) for phase *I* → phase *II* (i.e. the two phases merge as the system moves towards a point on the critical manifold), leading to

$$\boldsymbol{M_2}(i,j) = \begin{cases} \frac{\delta_{ij}}{c_j} + 2B_{ij} & for\ i < N \\ -S_{c,1j}{}^3\left\{Det\ \boldsymbol{M}_1^{(j,j)}\right\}\frac{S_{c,j1}{}^3}{c_{j,c}{}^2} & for\ i = N \end{cases} \qquad \text{with } i,j = 1,2,\ldots,N \tag{25}$$

with $\delta_{ij}$ the Kronecker delta. Analogous to the case for the spinodal manifold, an expression for the coordinates $c_{i,c}$ of the critical manifold ('*c*' stands for critical (manifold)) is conjectured, based on Eq (11) and earlier results for *N*=2 and *N*=3, as

$$c_{i,c} = \frac{1}{2\left(\sum_{j=1}^{N} B_{ij} S_{c,ji}\right)} \qquad \text{with } i = 1,2,\dots,N \tag{26}$$

It can be seen that the structure of the expressions for the coordinates $c_{i,c}$ (Eq (26)), $c_{i,sp}$ (Eq (20)), and $c_{i,s}$ (Eq (11)) is identical and the difference comes from the requirements for the associated parameters $S_{c,i1}$, $S_{sp,i1}$ and $S_{m,i1}$ (Eq (29) for the critical manifold and Eqs (8) and (12) for the manifold defined by the $c_{i,s}$). However, the role of the $c_{i,s}$ differs from that of the $c_{i,sp}$ and $c_{i,c}$ in the sense that the latter two represent real objects in the phase diagram, whereas the former only represent auxiliary coordinates (on the spinodal manifold) that allow us to calculate the coordinates $c_i^I$ and $c_i^{II}$ on the binodal manifold. Note that the non-negativity of coordinates $c_{i,c}$ for a specific choice of $B_{ij}$ and $S_{c,ij}$ in Eq (26) needs to be checked separately. The substitution of Eq (26) in Eq (25), leads to

$$\begin{aligned}
Det\ \boldsymbol{M_2} &= \sum_{j=1}^{N} (-1)^{j+N} \left\{\frac{\partial}{\partial c_j}(Det\ \boldsymbol{M_1})\right\} Det\ \boldsymbol{M}_1^{(j,N)} \\
&= \sum_{j=1}^{N} (-1)^{j+N+1} {S_{c,1j}}^3 \left\{Det\ \boldsymbol{M}_1^{(j,j)}\right\} \left\{Det\ \boldsymbol{M}_1^{(j,N)}\right\} \frac{{S_{c,j1}}^3}{{c_{j,c}}^2} \\
&= \sum_{j=1}^{N} -{S_{c,1N}}^3 \left\{Det\ \boldsymbol{M}_1^{(N,N)}\right\} \left\{Det\ \boldsymbol{M}_1^{(N,N)}\right\} \frac{{S_{c,j1}}^3}{{c_{j,c}}^2} \\
&= -{S_{c,1N}}^3 \left\{Det\ \boldsymbol{M}_1^{(N,N)}\right\}^2 \sum_{j=1}^{N} \frac{{S_{c,j1}}^3}{{c_{j,c}}^2} = 0
\end{aligned} \tag{27}$$

where use was made of a relation between minors of the determinant of the Hessian matrix $\boldsymbol{M_1}$ (Eq (28), see Appendix C).

$$Det\ \boldsymbol{M}_1^{(i,j)} = (-1)^{i+j} S_{sp,iN} S_{sp,jN} \left\{Det\ \boldsymbol{M}_1^{(N,N)}\right\} \tag{28}$$

where $\boldsymbol{M}_1^{(i,j)}$ is the minor of $\boldsymbol{M_1}$ relative to the element (*i*,*j*). As a result, $Det\ \boldsymbol{M_1} = 0 \wedge Det\ \boldsymbol{M_2} = 0$, provided that

$$\sum_{j=1}^{N} \frac{{S_{c,j1}}^3}{{c_{j,c}}^2} = 0 \tag{29}$$

It is noted that Eq (29) states an extra requirement for the critical points compared to the requirements for the points defining the spinodal manifold (see Eq (26) and Table 1). In other words, all points on the critical manifold lie on the spinodal manifold, but not the other way around. It is possible to substitute Eq (26) in Eq (29), but the present expression is simpler and therefore more insightful. The combination of Eqs (26) and (29) represents an ($N$-2) dimensional manifold (a critical point for $N$=2, a critical curve for $N$=3, etc). As expected, Eqs (26) and (29) reduce to the previously obtained results for binary [23] and ternary mixtures [25] (See Table 1). Again, it is noted that the existence of solutions for the critical point(s) for negative concentration coordinates is a property of the model (cf. Fig. 1 in Ref 21).

## Other generalizations of results for binary and ternary mixtures.

Finally, it is possible to generalise a number of results obtained previously for ternary mixtures [25]. To obtain $N$ phases, the phase separation criterion $\frac{{B_{ij}}^2}{B_{ii}B_{jj}} > 1$ should be satisfied for ($N$-1) binary combinations. In the case that $N$ phases segregate, one can conclude from the behavior of binary mixtures at high concentrations that the $N$ co-existing phases are located in the hyperplane defined by

$$\sum_{i=1}^{N} \sqrt{B_{ii}} c_i = c_0 \tag{30}$$

with $c_1, c_2, \ldots, c_N \rightarrow \infty$, leading to $c_0 \gg 0$ ($\mathrm{mol}^{1/2} \cdot \mathrm{m}^{-1/2}$) (cf. Eq (55) in Ref 25). Note that this expression does not contain the cross-virial coefficients $B_{ij}$, because in this limit each of the $N$ phases is composed of essentially pure components 1, 2, …, $N$.

Another straightforward generalization of results obtained for binary and ternary mixtures (cf. Eq (11) in Ref 24 and Eq (56) in Ref 25) for $P$=2 yields

$$\sum_{i=1}^{N} (c_i^I - c_i^{II}) + \frac{1}{2}(c_i^I + c_i^{II}) \ln\left(\frac{c_i^{II}}{c_i^I}\right) = 0 \tag{31}$$

using Eqs (12) and (13). All solutions to the co-existence equation fulfil Eq (31) or the generalizations thereof for higher $P$. For every additional phase, an additional equation is added of the form

$$\sum_{i=1}^{N} (c_i^I - c_i^{K}) + \frac{1}{2}(c_i^I + c_i^{K}) \ln\left(\frac{c_i^{K}}{c_i^I}\right) = 0 \qquad \text{with } K = III, IV, \ldots, P \tag{32}$$

Combining Eqs (31) and (32) results in similar expressions for any pair of phases. The remarkable aspect of Eqs (31)-(32) is that they do not contain any of the virial coefficients. A combination of concentration coordinates that does not fulfil Eqs (31)-(32) will not be a solution to the co-existence equations for *any* combination of virial coefficients.

# Results: examples

## A simple example for $N$=3 and $P$=2.

Because it is not customary to represent Eqs (1)-(5) and their solutions in terms of the parameters $S_{z,ji}$ (where $z$ can represent '$sp$', '$c$' or '$m$', i.e. the parameters associated with the spinodal manifold, the critical manifold, and the coordinates relevant to the tie-lines, respectively), an explicit example will-be discussed here. For $N$=3 there are two independent parameters $S_{z,ji}$. The example will address a case for $N$=3 to facilitate visualization, but the same principles also apply to larger numbers of components. Figure 1 displays potential combinations of $S_{z,21}$ and $S_{z,31}$ for a given set of virial coefficients for $N$=3 for a typical combination of values for the second virial coefficients that characterise polymer mixtures [32]. Note that these calculations go beyond those in Ref 25, which were restricted to 'symmetric' mixtures, having equal pure second virial coefficients $B_{ii}$ and equal cross second virial coefficients $B_{ij}$. The black dotted regions correspond to combinations of parameters $S_{z,ji}$ that give rise to physical solutions (i.e. non-negative concentrations for all coordinates $c_{i,z}$)

$$\frac{1}{c_{i,z}} = \sum_{j=1}^{N} B_{ij} S_{z,ji} \geq 0 \qquad \text{with } i = 1,2,\dots,N \tag{33}$$

where the rules from Eqs (8), (21) and (24) apply, which interrelate the parameters $S_{z,ji}$. Eq (33) represents a set of linear inequalities defining a linear programming problem without an optimalisation objective [33]. The white area corresponds to regions in which at least one coordinate is negative, and which should therefore be considered an unphysical solution. The boundaries between the dotted and white areas are given by the set of mangenta dotted lines $\sum_{j=1}^{N} B_{ij} S_{z,ji} = 0$ (one line for every $i$=*1,2,…,N*) and the $S_{z,j1} = 0$ axes with $j$=*1,2,…,N*), where $N$=3 in Figure 1. Green solid circles along the axes where either $S_{c,21}$ or $S_{c,31}$ are zero reflect (both the physical *and* unphysical) solutions associated with the critical points for $N$=2, when Eq (29) reduces to Eq (3) in Ref 24. The solid green curves in Figure 1 correspond to the combinations of $S_{c,21}$ and $S_{c,31}$ that fulfill Eq (29), and the part of the solid curves in the black dotted areas represent the physical solutions for the critical curve, Eq (29).

Figure 1 illustrates that for $N$=3 the regions of $(S_{z,21}, S_{z,31})$ parameter space that make up the physical solutions are not necessarily interconnected, but the segmentation in physical and unphysical solutions follows straightforward rules. Substitution of any combination of $(S_{sp,21}, S_{sp,31})$ from the dotted areas in Eq (20) will lead to a valid spinodal point. Similarly, any combination of $(S_{c,21}, S_{c,31})$ from the dotted area that satisfies Eq (29) will lead to a valid point on the critical curve. This principle will still apply for higher values of $N$, although keeping track of the rules that distinguish physical and unphysical solutions becomes more complicated.

Although the information in Figure 1 is conceptually sufficient to calculate the spinodal and critical curve, the need to evaluate the $(S_{z,21}, S_{z,31})$ parameter space in the range <-∞,∞> for either parameter makes this a less practical approach. Fortunately, the relations between the parameters in Eqs (8), (21) and (24) are of help here, and can be used to transform the parameter space in Figure 1 by splitting it in three parameter subspaces consisting of the positive quadrants of $(S_{z,21}, S_{z,31})$, $(S_{z,12}, S_{z,32})$ and $(S_{z,13}, S_{z,23})$, as is illustrated in Figure 2. The physical solutions to the problem (black dotted area) can be found in a narrow strip in each of the quadrants (grey and black dotted area), which allows for a full but straightforward numerical evaluation. Note that Figure 2 is slightly more complicated than might appear at first sight as the axes may reflect either $S_{z,ij}$ or its inverse $S_{z,ji}$, depending on the quadrant that is being considered. The approach described here is also the first step towards the calculation of the binodal, which is not discussed here as it involves a numerically more complex procedure.

The identified combinations of parameters that give rise to physical solutions in Figure 2 can be used to construct the spinodal and the critical curve in a rather straightforward way by substituting them in Equ. (20). The result is shown in Figure 3, with the same diagram being shown from two different points of view. To overcome the difficulty of showing a three-dimensional shape in a two-dimensional image, a video showing the diagram from multiple directions is provided in the Supplementary Information as well. Note that the criterion used to calculate the spinodal includes solutions that separate instable areas, as for example in Fig 2f in Ref 25 (constricting areas characterised by one negative eigenvalue of the Hessian matrix $\boldsymbol{M_1}$ versus areas characterised by two negative eigenvalues).

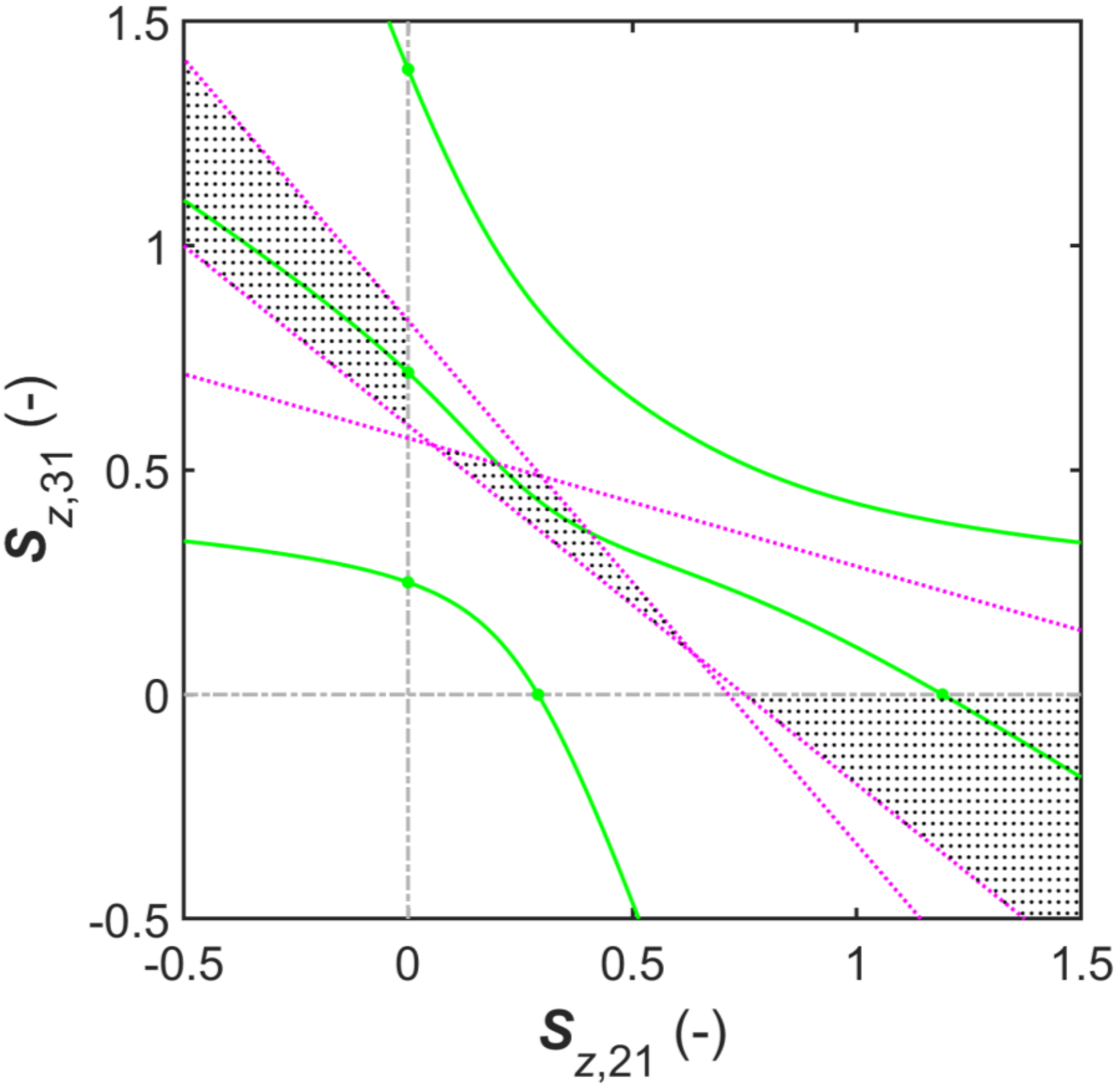

Figure 1. Map of part of the $(S_{z,21}, S_{z,31})$ parameter space (with *z*='*sp*', '*c*' or '*m*'). The area indicated by (░) depicts physically relevant solutions, i.e. positive coordinates for the critical curve or the spinodal surface (or the coordinates in Eq (11)). The area indicated by white regions depict unphysical parts of parameter space where not all three coordinates $c_{i,z}$ (with *i*=1,2,3) are positive; The lines indicated by the mangenta dotted line depict lines for which $1/c_{1,z} = 0$, $1/c_{2,z} = 0$ and $1/c_{3,z} = 0$ (cf. Eq (33)), which segment the parameter space in physically relevant and unphysical regions; The (green solid lines) depict relations between $S_{c,21}$ and $S_{c,31}$ according to Eq (29); and the points indicated by green solid circle depict both physical *and* unphysical solutions for the critical points for binary mixtures (for $S_{c,21} = 0.290, 1.191, 6.519$ (not shown) or $S_{c,31} = 0.250, 0.718, 1.394$), which are located on the grey dash-dotted line that represents either $S_{c,31} = 0$ or $S_{c,21} = 0$. Calculations have been performed for *N*=3 with virial coefficients $B_{11}$=1.5, $B_{22}$=1, $B_{33}$=3, $B_{12}$=2, $B_{13}$=2.5, $B_{23}$=3.5 ($m^3$/mol).

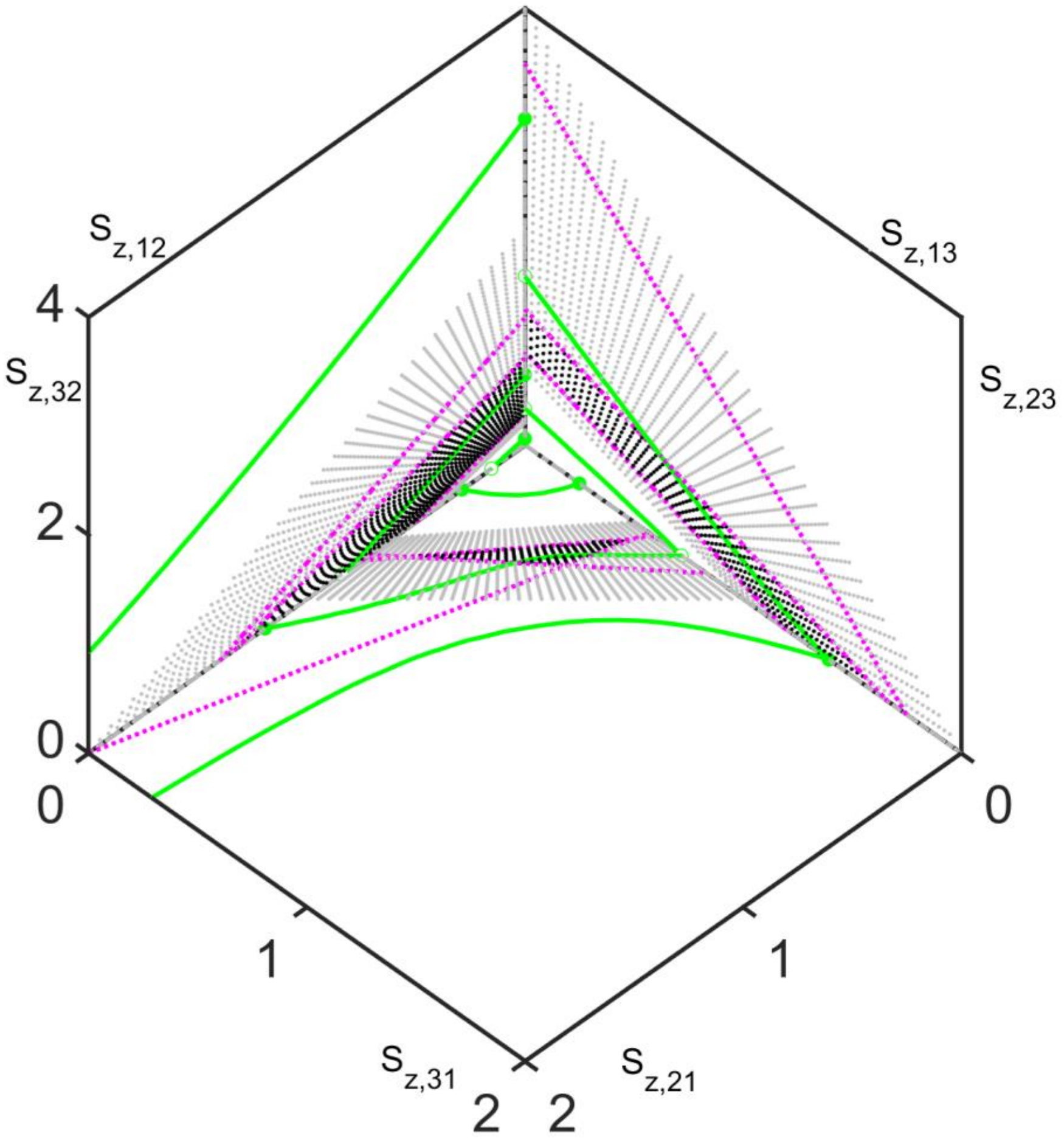

Figure 2: Same information as plotted in Figure 1, but now shown for the positive quadrants of $(S_{z,21}, S_{z,31})$, $(S_{z,12}, S_{z,32})$ and $(S_{z,13}, S_{z,23})$ (front-bottom, back-left, back-right, respectively). Note that $S_{z,ij} = 1/S_{z,ji}$. Physical (black dotted area) and unphysical (grey dotted area) solutions for the parameter space from which the spinodal surface (cf. Eq (20)) and critical curve (cf. Eq (26) and Eq (29)) can be evaluated. The area outside the strip that is numerically evaluated (white) represents unphysical solutions too. Dotted mangenta lines represent the curves for $1/c_{1,z} = 0$, $1/c_{2,z} = 0$ and $1/c_{3,z} = 0$ (cf. Eq (33)). Solid green curves reflect the solutions for the parameters satisfying Eq (29) that can be used to determine the critical curve. Calculations were performed for $N$=3 with virial coefficients $B_{11}$=1.5, $B_{22}$=1, $B_{33}$=3, $B_{12}$=2, $B_{13}$=2.5, $B_{23}$=3.5 ($m^3$/mol).

(a)

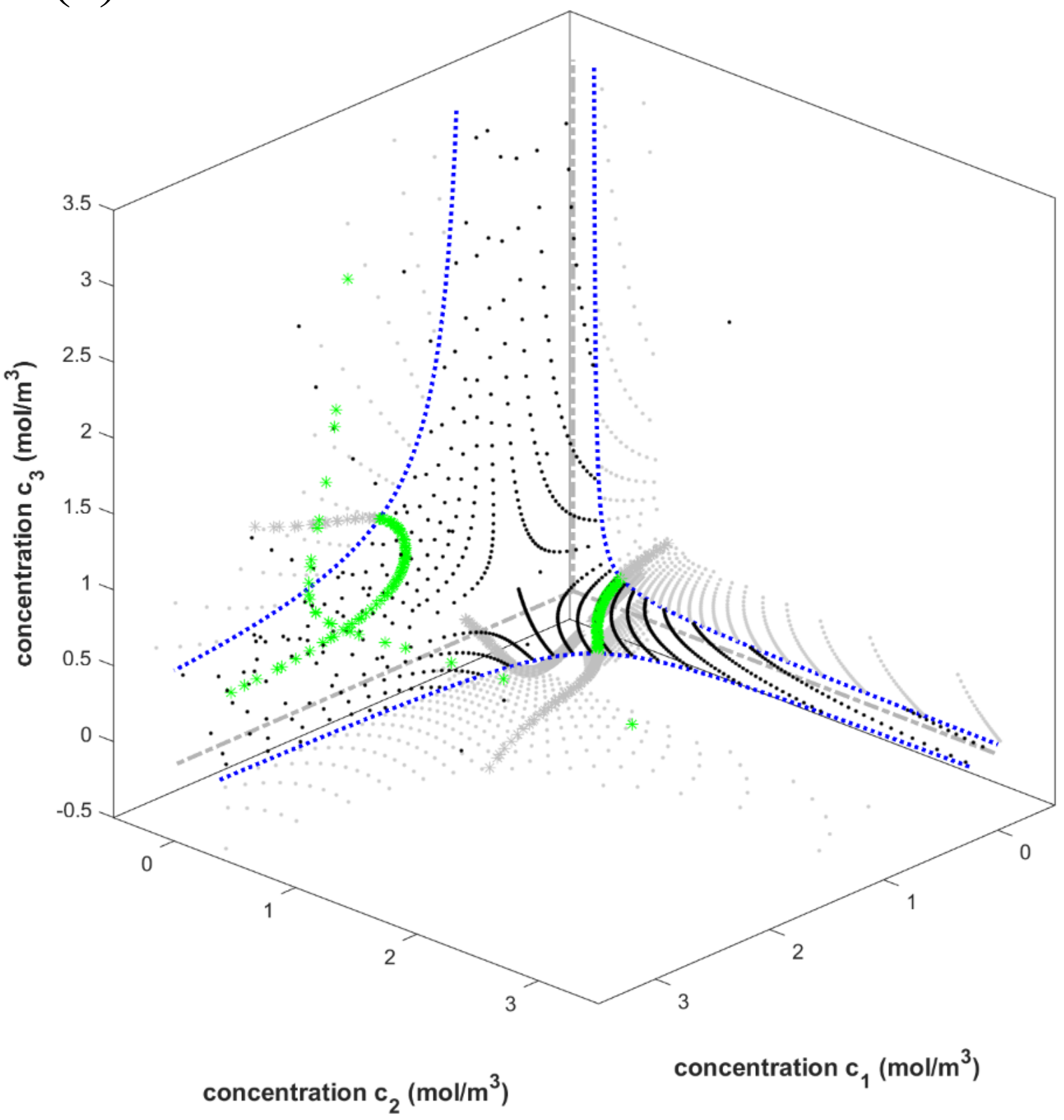

(b)

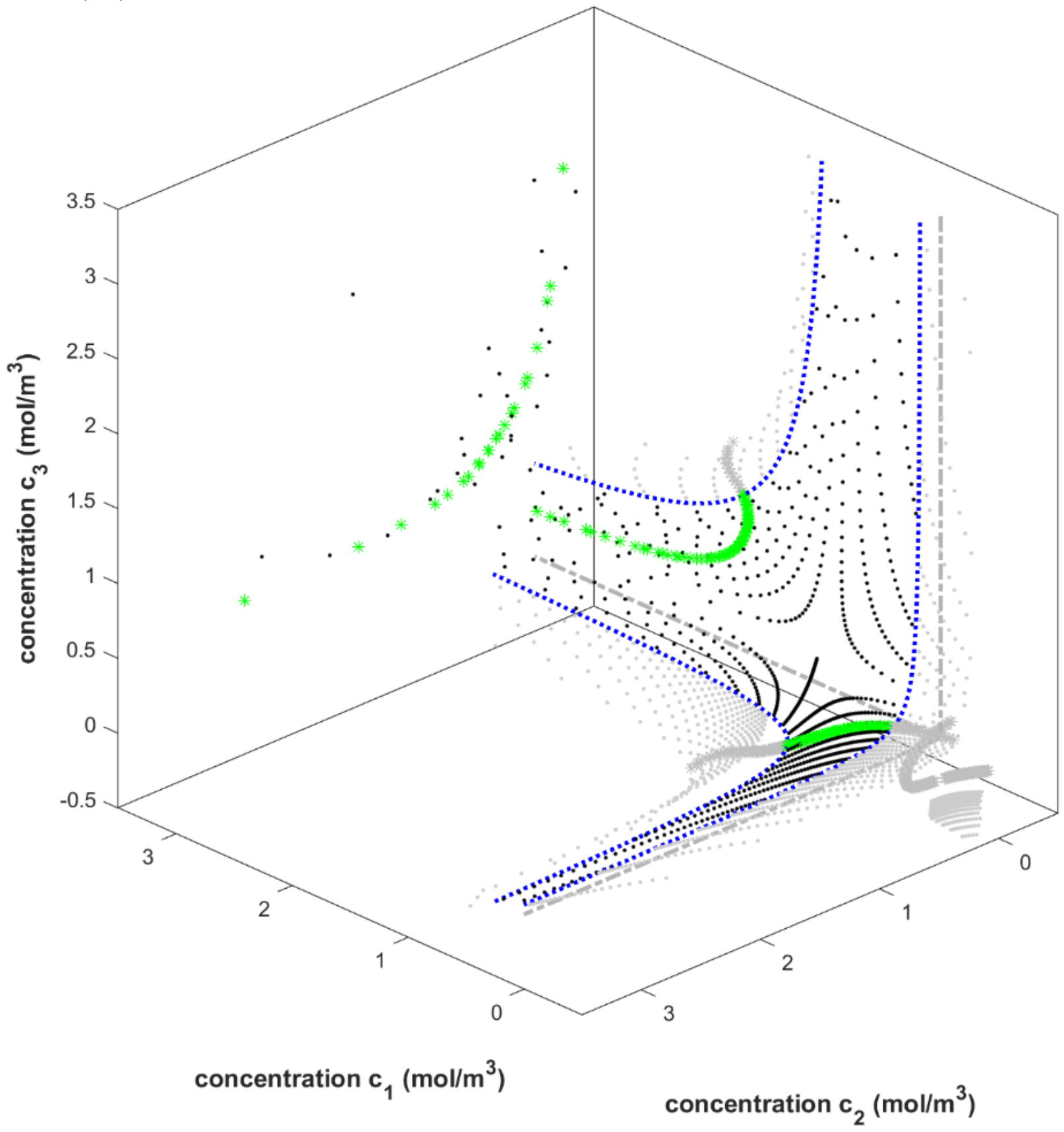

Figure 3: (a,b) Spinodal surface and critical curve for the same parameter choice as in Figure 2 ($N$=3 with virial coefficients $B_{11}$=1.5, $B_{22}$=1, $B_{33}$=3, $B_{12}$=2, $B_{13}$=2.5, $B_{23}$=3.5 ($m^3$/mol)), shown from two different directions. Features plotted in black refer to physical solutions, features plotted in grey to unphysical solutions (at least one of the concentration coordinates is negative). Symbols: (.) spinodal surface; (green asterisks) critical curve; (blue dotted lines) spinodal curve in the binary planes; (-.-.) axes ($c_1$,$c_2$)=(0,0) or ($c_2$,$c_3$)=(0,0) or ($c_3$,$c_1$)=(0,0). The spinodal surface and critical curve in the left-top hand of panel (b) separates two unstable regions of the phase diagram (comparable to what is shown in Fig 2f in Ref 25). A video that reflects the three-dimensional aspects of this diagram more clearly is available in the Supplementary Information.

## A complex example.

Although the prime focus of the present paper is for $N \lesssim 10^3$, the equations are also valid for $N \gtrsim 10^3$. Since RMT is valid typically for $N \gtrsim 10^3$, it is of interest to make a comparison between both approaches. One should keep in mind that RMT makes some assumptions on the statistical properties of the second virial coefficients, which is not the case for the approach described in the present work.

A recent paper by Thewes et al. will be used as a starting point for discussing the results and assumptions related to RMT [26]. This RMT approach allows for the calculation of the limit of stability in mixtures of many components using (only) averages and standard deviations of the second virial coefficients of the components. The assumptions on the statistical properties of the second virial coefficients simplify the calculation of the limit of stability considerably. Table 2 shows the relation between the various parameters in the paper of Thewes et al. and the parameters used in the present work.

| **Description** | **Thewes et al. [26]** | **Present work** |
|---|---|---|
| Number of components | $M$ | $N$ |
| Virial coefficient | $\epsilon_{\alpha\gamma}$ | $2TB_{ij}$ |
| Average virial coefficient | $b = -\langle\epsilon_{\alpha\gamma}\rangle$ | $-2T\langle B_{ij}\rangle$ |
| Variance of virial coefficient ($s^2$) | $s$ | |
| Normalised variance of the virial coefficients | $\tilde{s} = \frac{s}{\sqrt{M}}$ | |
| Total concentration | $\rho = \sum_{\alpha=1}^{M} \rho_\alpha$ | $c_{tot} = \sum_{i=1}^{N} c_i$ |
| Average concentration | $\bar{\rho} = \frac{\rho}{M}$ | $\bar{c} = \frac{c_{tot}}{N}$ |
| Relative concentration | $y_\alpha = \frac{\rho_\alpha}{\bar{\rho}}$ | $\frac{c_i}{\bar{c}}$ |
| Solvent concentration | $\rho_0$ | |
| Parameter, slope of tie-line in the plane ($i$,$j$) | | $-S_{m,ij}$ |
| Parameter, slope of critical manifold in the plane ($i$,$j$) | | $-S_{c,ij}$ |
| Parameter to calculate the spinodal | | $-S_{sp,ij}$ |

Table 2: Relations between the variables used in Ref 26 and the present paper. Despite playing the same role, the definition of the second virial coefficients ($\epsilon_{\alpha\gamma}, B_{ij}$) differs between both papers.

Thewes et al. [26] do not distinguish the pure second virial coefficients from the cross second virial coefficients when assigning values to them, since they draw them from a single assumed normal distribution, but they do respect the symmetry requirement $\epsilon_{\alpha\gamma} = \epsilon_{\gamma\alpha}$. The solvent molecules are taken into account explicitly, while a cross virial coefficient of zero is assigned to the interactions of the solvent with the other components in the mixture. In the present paper this could be achieved by replacing the equation for the osmotic pressure by one for the chemical potential of the solvent [23].

Thewes et al. distinguish between two main types of phase separation: condensation (C) referring to changes in concentration which are similar for all components (in the present paper: all $S_{m,ij} < 0$), and demixing (D) referring to situations where some components are enriched in one phase and depleted in other phase(s) (in the present paper: many $S_{m,ij} > 0$ occur). Demixing is further sub-divided in random demixing (RD), localized demixing and composition-driven demixing (CD).

Next, Thewes et al. consider four regions in the phase diagram and evaluate the character of phase separation in those regions: (1) equal concentration of all components (previously investigated already by Sear and Cuesta [16]), (2a) one component dominates the mixture, (2b) two components dominate the mixture (in their Supplementary Material section), and (3) a beta distribution for the concentrations of the components.

Condensation happens mostly at low concentrations, while random and composition-driven demixing happen at higher concentrations. Random demixing tends to occur in scenarios where all components are present in the same amount, whereas composition-driven demixing is more common for situations in which the concentration of one of the components dominates. In the localised demixing case just a few components dominate the phase behavior.

It is not possible to visualize the *N*-component case addressed by Thewes et al. [26], but it would be useful to identify the above regions of parameter space in terms of the formalism in the present paper. For this purpose, a simpler *N*=3 version of the special cases (1), (2a) and (2b) in the work by Thewes et al. is considered. This *N*=3 version is outside the scope of the work by Thewes et al. [26].

The case of equal concentration of all components (Example 1 in Ref 26, previously investigated by Sear and Cuesta [16]) requires

$$c_{i,sp} = c_{j,sp} = c_{tot}/N \qquad \text{for all } i,j = 1,2,\dots,N \tag{34}$$

corresponding to

$$\frac{1}{2\left(\sum_{k=1}^{N} B_{ik} S_{sp,ki}\right)} = \frac{1}{2\left(\sum_{k=1}^{N} B_{jk} S_{sp,kj}\right)} \tag{35}$$

This condition is represented by the blue dot in Figure 4.

The case in which one of the components, e.g. $c_{i,sp}$, dominates the mixture (Example 2 in Ref 26) requires that the denominator in Equation (36)

$$c_{i,sp} = \frac{1}{2\left(\sum_{k=1}^{N} B_{ik} S_{sp,ki}\right)} \qquad (36)$$

is much smaller compared to the concentrations of the other components for this point on the spinodal surface. In our parameter space this corresponds to parameter combinations that are found close to the edge of the physical region (black dotted area) in Figure 1 or Figure 2. This edge is indicated qualitatively by red lines in Figure 4. In a similar way, two dominant components (Supplementary Material to their paper) can be found close to the edge of the physical region where two lines described by $\sum_{k=1}^{N} B_{ik} S_{sp,ki} = 0$ intersect. The relevant locations are indicated qualitatively by three green dots in Figure 4.

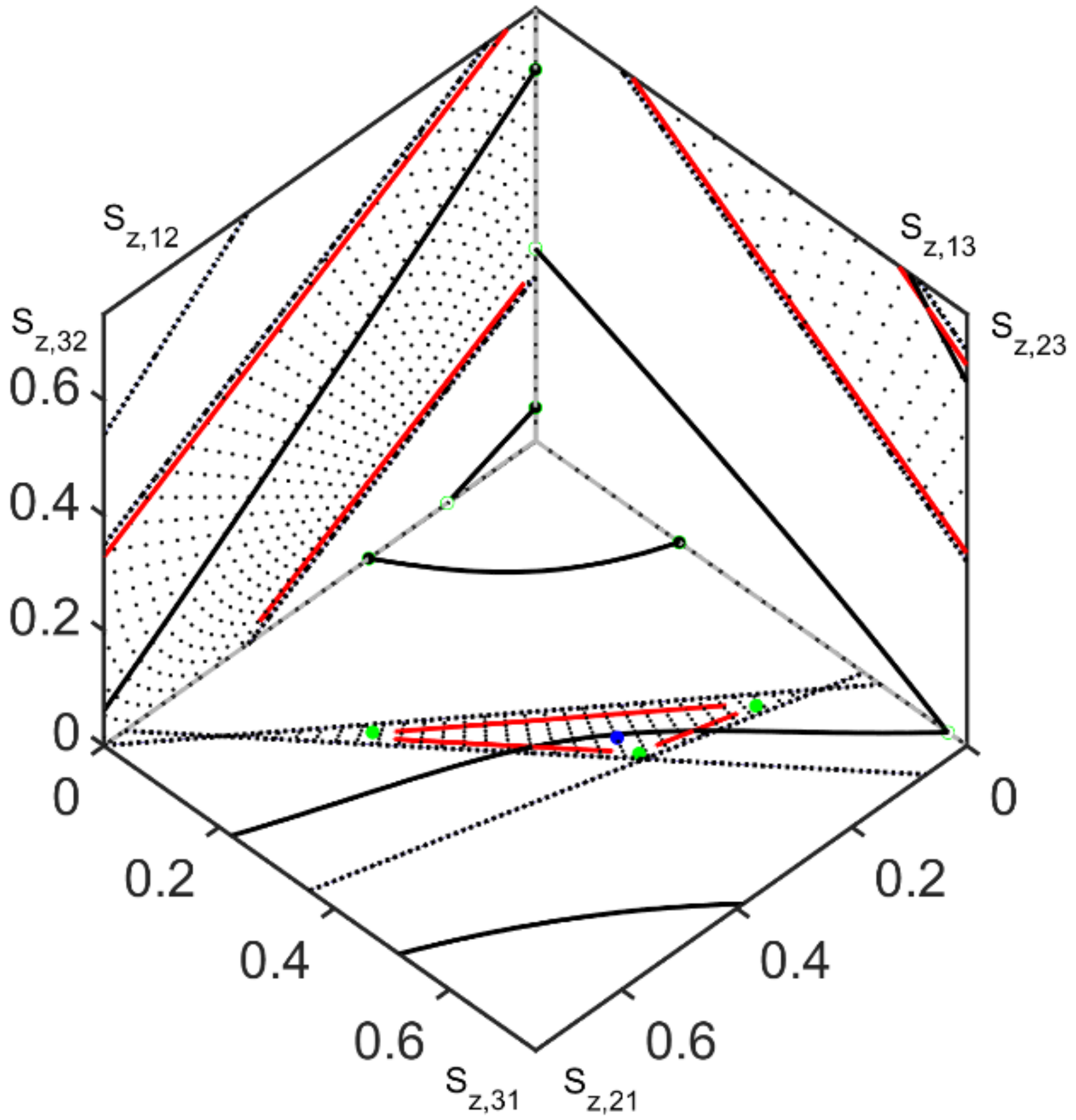

Figure 4: Zoomed-in version of Figure 2, qualitatively indicating the parameter regions for the examples discussed in the paper of Thewes et al. [26], but translated to the of $N$=3 case to facilitate visualisation: (blue dot) Equal concentrations of all components as in their Example 1; (red lines) One dominant species as in their Example 2 - close to the edges for the 'physical' regions, but away from intersections with other edges; (green dots) Two dominant species as in an Example described in their Supplementary Material - at the intersection of two edges. Black dotted regions indicate physically relevant parameter choices. The remaining lines have the same interpretation as in Figure 2. The second virial coefficients were chosen as $B_{11}$=1.5, $B_{22}$=1, $B_{33}$=3, $B_{12}$=2, $B_{13}$=2.5, $B_{23}$=3.5 ($m^3$/mol).

## On the applicability window for the present approach.

The results in the present work are proposed to complement the RMT approach, in case one is interested in the spinodal manifolds. Again, the coexistence regions are not adrresed in RMT, but are considered in our approach. In the limit $N \to \infty$, RMT is exact (under the assumption of statistical independence of the virial coefficients and some other assumptions underlying this theory) and likely more efficient in calculating the spinodal when compared to the new results in the present work – despite the fact that these results are exact as well. On the other hand, RMT is expected to fail for smaller *N*. This opens an applicability window for the results of the present work.

Next the question is investigated how small *N* needs to be to have deviations from the exact result for $N \to \infty$. RMT calculates the eigenvalues of the Hessian matrix $\boldsymbol{M}_1$ and determines the conditions under which the smallest of its eigenvalues turns to zero, reflecting the boundary of the spinodal area. One key property of RMT for large *N* is that the distribution of eigenvalues of matrix $\boldsymbol{M}_1$ follows the so-called Wigner's semi-circle law, with single outliers for lower and higher values in case the average of the virial coefficients in $\boldsymbol{M}_1$ is lower or higher than zero (In the case of Example 1 in the previous section, the requirement that $c_{i,sp} = c_{tot}/N$ for all $i$ only leads to a shift in the overall eigenvalue spectrum of $\boldsymbol{M}_1$ compared to the eigenvalue spectrum of the matrix of second virial coefficients $\boldsymbol{B}$, with $B(i,j) = B_{ij}$) [16]. Deviations from the semi-circle law will occur at finite *N*, and the present section is dedicated to find a qualitative estimate for which *N* the semi-circle law can still be considered a reasonable approximation to the actual distribution. For that, the eigenvalue distribution for a given set of virial coefficients $B_{ij}$ was calculated, where the values of the virial coefficients were drawn from a normal distribution.

In Figure 5 a number of eigenvalue distributions are shown for different values of *N*, where the number of bins was chosen to be (the rounded value of) $\sqrt{N}$ to ensure a sufficient number of eigenvalues per bin. The average value for the virial coefficients is taken to be zero, and as a consequence no 'outliers' (i.e. small peaks left or right of the distribution) are observed. For $N$=10 and $N$=100, the distribution will depend strongly on the specific draw of the virial coefficients, but it can be shown that the semi-circle law is satisfied if one repeats the calculation thousands of times and adds each result to the distribution. At $N$=1000, the semi-circle law starts to become a fair representation of semi-circle distribution, although deviations between the actual distribution and the semi-circle law are still present. At $N$=10000, the semi-circle law gives a very accurate, although not exact, representation of the calculated distribution. The value of *N* at which the semi-circle law is not satisfied anymore depends of course on the specific criterion that is used to quantify the similarity, but for now an order-of-magnitude assessment suffices, indicating that this occurs for $N \approx 10^3$.

Thus, the results of the present work have a better validity than the RMT for a rather large range of from $N$=2 to $\sim 10^3$. In practice, the present theoretical approach can probably best be used for mixtures up to one or two dozen components.

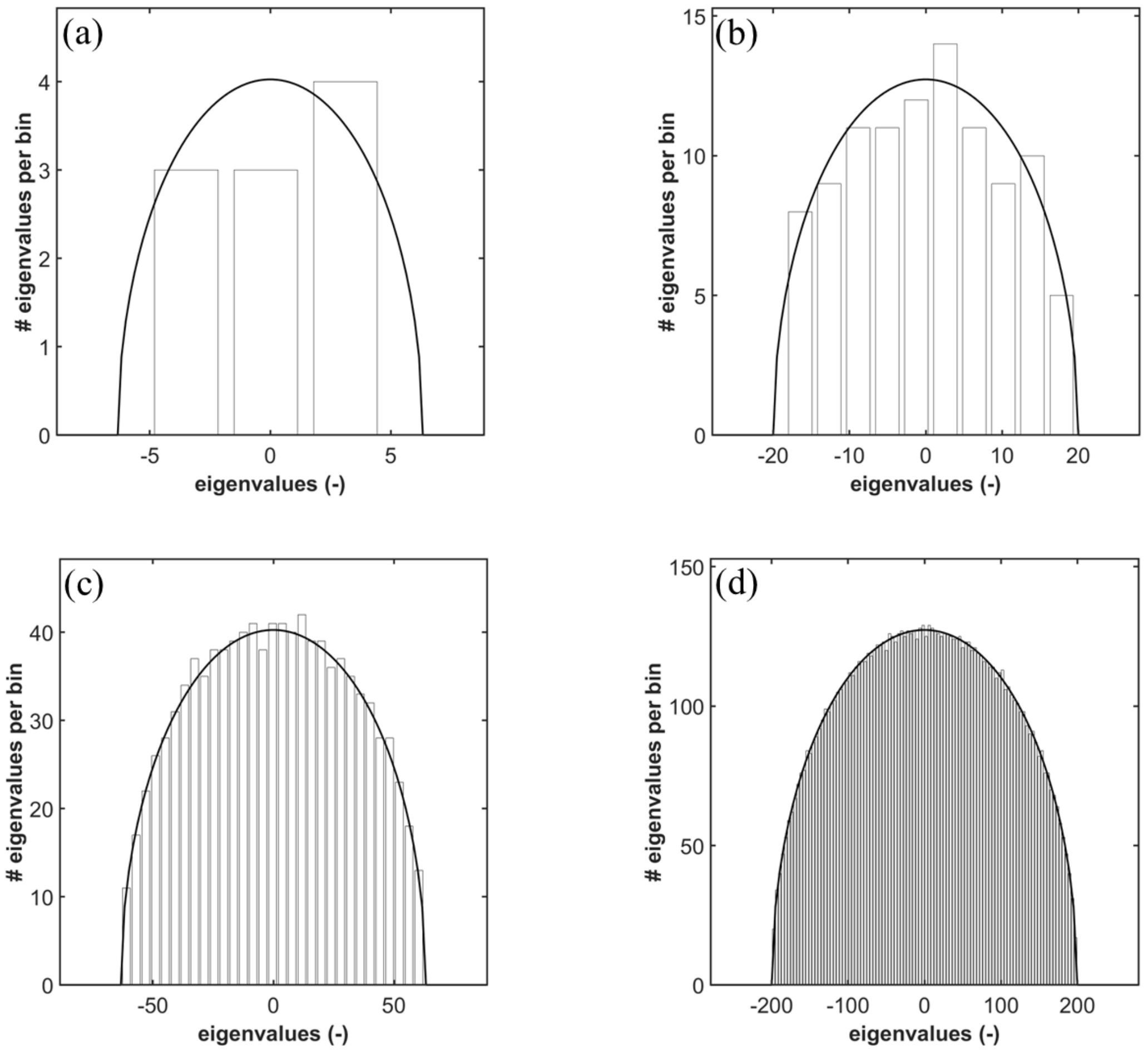

Figure 5: Eigenvalue spectrum of matrix $\boldsymbol{B}$ containing the virial coefficients $B_{ij}$, for $N$=10 (a), 100 (b), 1000 (c) and 10000 (d). The solid line represents the distribution according to Wigner's semi-circle law. In the case of Example 1 in the work of Thewes et al. [26], the *shape* of the eigenvalue distribution of $\boldsymbol{B}$ is the same as of $\boldsymbol{M}_1$, but only *shifted* relative to the eigenvalue distribution of $\boldsymbol{M}_1$. The virial coefficients in the calculation were drawn from a normal distribution with average 0 and variance 1. In cases for which the variance differs from unity, the horizontal axis of the eigenvalue spectrum represents eigenvalue/variance.

## Discussion

In this paper a generalization is presented of earlier work on binary and ternary mixtures to mixtures of $N$ macromolecular components, with applications in mind for typically $2 \leq N \lesssim 10^3$. The model can be used in cases where the second virial coefficients cannot be considered statistically independent. Another potential application is in calculations on polydisperse mixtures where the molecular weight distribution of the components is approximated by binning specific molecular weight ranges of the distribution as separate components ($N$ being dozens). Surprisingly, the model in Eq (1) can be reformulated in terms of the parameters characterizing the spinodal and critical manifolds and co-existing phases. This is not obvious at all when considering the governing Eq (1). In addition, the current approach allows to determine the critical manifolds and co-existence phases for small and large numbers of components.

The first question that could be posed is whether the expression for the Helmholtz free energy in Eq (1) is the most suitable for the challenge at hand. As an alternative, for example, the Flory-Huggins model could be considered. However, in Ref 23 a comparison between the Flory-Huggins model and the present model was made for *N*=2, and it was found that both models could be mapped onto each other up to second order of concentration. Both models can be considered as mean field theories and their usefulness follows from the insight that is provided, more than from exact predictions. However, the model used in Eq (1) is expected to still give fair predictions (see e.g. Ref 32 for examples for *N*=2). The main challenge in the application of these models to systems with many components comes from obtaining the proper model parameters, be it Flory-Huggins parameters or second virial coefficients. Random Matrix Theory gives direct guidance on how the statistics of the second virial coefficient distribution affects the predicted spinodal [16,26], which is an advantage if one wants to choose a relevant set of values without knowing each and every individual value.

In this paper, formal analytical expressions for spinodal points in *N* dimensions are obtained in terms of the second order virial coefficients and a set of parameters $S_{sp,ij}, S_{c,ij}$ and $S_{m,ij}$ subject to well-defined product rules (see Table 1). The expressions for binary [23] and ternary [25] mixtures were previously derived, but in the present paper these results are generalised to *N* component mixtures. The actual calculation involves a choice for the values of each of the *N*(*N*+1)/2 virial coefficients $B_{ij}$. This choice can either be based on the physical properties of the components and determined experimentally or estimated theoretically, or can alternatively be obtained through *a priori* assumptions on the statistical distribution of the virial coefficients [18]. The present paper provides a procedure to determine the spinodal and critical points using Eqs (20) and (26), where the calculation is achieved by choosing the $S_{c,ij}$ or $S_{sp,ij}$ in the range <-∞,∞> (although transformation of the parameter space strongly restricts these ranges, as was illustrated for *N*=3 and positive virial coefficients in the Results section). For physically relevant solutions, the choices of $S_{c,ij}$ or $S_{sp,ij}$ should lead to non-negative coordinates $c_{i,c}$ or $c_{i,sp}$ – something that can be confirmed easily by substituting the selected $S_{c,ij}$ or $S_{sp,ij}$ in Eq (20). For the calculation the critical points, additionally one of the $S_{c,ij}$ should be chosen in such a way that Eq (29) is fulfilled also (cf. Figure 1 for *N*=3). Assuming $S_{c,j1}$ is the remaining parameter to be chosen, the requirement from Eq (29) leads to an expression where the highest power in $S_{c,j1}$ is of the form ${S_{c,j1}}^3/{c_{j,c}}^2$ (with $1/c_{j,c}$ linear in $S_{c,j1}$) resulting in a fifth-order polynomial in this remaining parameter $S_{c,j1}$. This polynomial can be solved numerically.

One of the benefits of the present mathematical representation of Eqs (6) and (7) is that the signs of the chosen $S_{m,ij}$ makes immediately clear if a solution represents segregative or associative phase separation for the pair of components $i$ and $j$. Parameter choices with $S_{m,ij} > 0$ describe segregative phase separation between components, whereas choices with $S_{m,ij} < 0$ describe associative phase separation. As a consequence, both phenomena need to occur in phase diagrams for $N$≥3, as was earlier inferred for ternary mixtures ($N$=3) [25]. The necessary existence of associative phase separation ($S_{m,ij} < 0$) in some region in the phase diagram, even for systems with only positive virial coefficients $\left(B_{ij} > 0\right)$, follows from the observation that a valid and physical combination of two positive $S_{m,ik}$ and $S_{m,kj}$ can be combined to a negative $S_{m,ij} = -S_{m,ik}S_{m,kj} < 0$ (Note that Minton studied associative phase separation for $N$=2 for a slightly more complicated version of the present model [34]).

Having chosen the virial coefficients, it is possible to sketch an algorithm to calculate the phase diagrams. The first step is to choose sets of ($N$-1) values for the $S_{sp,ij}$. As indicated above, the choices of $S_{sp,ij}$ should lead to non-negative coordinates $c_{i,sp}$. This can be established *a posteriori*, but it would be more efficient if an algorithm were to be identified that ensured non-negative coordinates *a priori* as the fraction of composition space with all coordinates being positive reduces as $2^{-N}$ for $N$ components. For the $N$=3 case, an effective approach was demonstrated in the Results section, but for larger $N$ a more general algorithm is required. A possible strategy is to start from the $S_{c,ij}$ for a binary (sub-)mixture (obtaining one $S_{c,ij}$ by solving Eq (20) in Ref 23 and choosing the others zero, and systematically evaluating choices of $S_{sp,kl}$ (with ($k$,$l$)≠($i$,$j$)) around this initial choice for $S_{c,ij}$, and repeat this for other binary (sub-)mixtures in the $N$-component mixture. It should be noted that this strategy will not automatically provide *all* relevant solutions, as regions leading to non-negative coordinates in composition space are not necessarily interconnected in the parameter space defined by all $S_{sp,ij}$ (cf. Figure 1 for $N$=3). A next step could be to follow the hyperplanes defined by $1/c_{i,\mathrm{sp}} = 2\left(\sum_{j=1}^{N} B_{ij}S_{sp,ji}\right) = 0$ and explore the non-negativity of the coordinates around this hyperplane. Only parameter choices resulting in all coordinates $c_{i,\mathrm{sp}}$ being non-negative need to be retained here. The solutions derived this way circumvent the intermediate step of calculating of the eigenvalues of the Hessian matrix $\boldsymbol{M_1}$ that features in approaches based on Random Matrix Theory. In addition, the present approach is applicable to all $N$≥2, and there are no requirements on the statistical properties of the virial coefficients. It should be emphasized that the above procedure to calculate the spinodal is an example of the simplification that the method introduced in this paper has brought.

The second step is to choose sets of ($N$-2) values for the $S_{c,ij}$. This can be done by repeating the above steps, but taking Eq (29) into account. Possibly a more efficient algorithm is to take the parameter choices representing non-negative coordinates from the previous steps and replace one of the parameters to ensure that the set satisfies Eq (29). This procedure will still produce some parameter combinations representing negative coordinates, but likely at a much lower frequency than if one starts from a more random choice. Again, only parameter choices resulting in all coordinates $c_{i,c}$ being non-negative need to be retained here. Also, the calculation of the critical manifold has become much simpler using the expressions described in this paper.

The third step would start from the critical coordinates $c_{i,c}$ and parameters $S_{c,ij}$ identified in the previous steps, and use them as input for $S_{m,ij}$ (and therefore $c_{i,s}$) in the calculation of the tie-lines. This will require a multi-dimensional optimisation procedure to systematically vary some of the coordinates on the binodal and find consistent solutions for the parameters $S_{m,ij}$ (and therefore $c_{i,s}$) and the coordinates for the binodal. These solutions need to satisfy Eqs (8), (11), (14) and (16) simultaneously. Although the techniques described in the present paper provide more insight into the calculation of the binodal for the present model, it should still be considered a challenging numerical task.

Executing and optimising the above sketch of an algorithm is a daunting task and outside the scope of the present paper, which had the aim to identify a possible route to do such calculations for many components. If attempted, it is probably wise to restrict oneself to a small part of the phase diagram (e.g. the calculation of part of the spinodal), as it was found previously that doing these calculations for the full phase diagram and for a general choice of virial coefficients is not straightforward, even for $N$=3 (where the calculations in Ref 25 were restricted to 'symmetric' mixtures, with equal pure second virial coefficients $B_{ii}$ and equal cross second virial coefficients $B_{ij}$).

Finally, it is noted that the phase separation criterion ${B_{ij}}^2 > B_{ii}B_{jj}$ allows for negative second cross-virial coefficients $B_{ij}$. The critical point coordinates $c_{i,c}$ are non-negative when $1/c_{i,c} = \sum_{j=1}^{N} B_{ij}S_{c,ji} \geq 0$ for all $i$ (cf. Eq (33) and where $B_{ij}$ is a symmetric matrix, which is always satisfied when both the virial coefficients and the tangents have the same sign and sometimes satisfied if they have opposite signs. This allows for the description of both segregative ($S_{m,ij} > 0$) and associative ($S_{m,ij} < 0$) phase separation for $N$≥3.

# Appendices

## Appendix A: Matrix elements for the co-existence matrix

The derivation for the expressions of the elements of the co-existence matrix in Eqs (16) and (17) starts from Eq (12) for the osmotic pressure and Eq (15) for the slopes of the tie-lines. The aim is to eliminate all but one of the $c_j^I$ for each of the rows in the matrix equation. This can be achieved by rewriting the expression for the slope as

$$(\frac{c_j^I}{c_{j,s}} - 1) = (\frac{c_j^{II}}{c_{j,s}} - 1) \; - S_{m,ji}\frac{c_{i,s}}{c_{j,s}}\left((\frac{c_i^I}{c_{i,s}} - 1) - (\frac{c_i^{II}}{c_{i,s}} - 1)\right) \tag{37}$$

and rewriting the osmotic pressure equation as

$$S_{m,i1}\left(\left(\frac{c_i^I}{c_{i,s}} - 1\right) + \left(\frac{c_i^{II}}{c_{i,s}} - 1\right)\right) + \sum_{\substack{j=1 \\ j\neq i}}^{N} S_{m,j1}\left(\left(\frac{c_j^I}{c_{j,s}} - 1\right) + \left(\frac{c_j^{II}}{c_{j,s}} - 1\right)\right) = 0 \tag{38}$$

Substitution of Eq (37) in Eq (38) leads to

$$\begin{aligned} &S_{m,i1}\left(\left(\frac{c_i^I}{c_{i,s}} - 1\right) + \left(\frac{c_i^{II}}{c_{i,s}} - 1\right)\right) \\ &\quad + \sum_{\substack{j=1 \\ j\neq i}}^{N} S_{m,j1}\left(2\left(\frac{c_j^{II}}{c_{j,s}} - 1\right) - S_{m,ji}\frac{c_{i,s}}{c_{j,s}}\left(\frac{c_i^I}{c_{i,s}} - 1\right) + S_{m,ji}\frac{c_{i,s}}{c_{j,s}}(\frac{c_i^{II}}{c_{i,s}} - 1)\right) = 0 \end{aligned} \tag{39}$$

Rearranging Eq (39)

$$\begin{aligned} &\left(S_{m,i1} + \sum_{\substack{j=1 \\ j\neq i}}^{N}\left(-S_{m,j1}S_{m,ji}\frac{c_{i,s}}{c_{j,s}}\right)\right)\left(\frac{c_i^I}{c_{i,s}} - 1\right) + \left(S_{m,i1} + \sum_{\substack{j=1 \\ j\neq i}}^{N}\left(S_{m,j1}S_{m,ji}\frac{c_{i,s}}{c_{j,s}}\right)\right)\left(\frac{c_i^{II}}{c_{i,s}} - 1\right) \\ &\quad + \sum_{\substack{j=1 \\ j\neq i}}^{N} S_{m,j1}\left(2\left(\frac{c_j^{II}}{c_{j,s}} - 1\right)\right) = 0 \end{aligned} \tag{40}$$

Multiplication by $S_{m,i1}/c_{i,s}$ leads to

$$\left(\frac{S_{m,i1}^2}{c_{i,s}}+\sum_{\substack{j=1\\j\neq i}}^{N}\left(\frac{S_{m,j1}^2}{c_{j,s}}\right)\right)\left(\frac{c_i^I}{c_{i,s}}-1\right)+\left(\frac{S_{m,i1}^2}{c_{i,s}}+\sum_{\substack{j=1\\j\neq i}}^{N}\left(-\frac{S_{m,j1}S_{m,j1}}{c_{j,s}}\right)\right)\left(\frac{c_i^{II}}{c_{i,s}}-1\right) + \sum_{\substack{j=1\\j\neq i}}^{N}\frac{2S_{m,i1}S_{m,j1}}{c_{i,s}}\left(\frac{c_j^{II}}{c_{j,s}}-1\right)=0 \tag{41}$$

which can be rearranged as

$$\left(\frac{c_i^I}{c_{i,s}}-1\right)=\frac{\left(\sum_{\substack{j=1\\j\neq i}}^{N}\left(\frac{S_{m,j1}^2}{c_{j,s}}\right)-\frac{S_{m,i1}^2}{c_{i,s}}\right)}{\left(\sum_{j=1}^{N}\left(\frac{S_{m,j1}^2}{c_{j,s}}\right)\right)}\left(\frac{c_i^{II}}{c_{i,s}}-1\right)-\frac{\sum_{\substack{j=1\\j\neq i}}^{N}\frac{2S_{m,i1}S_{m,j1}}{c_{i,s}}\left(\frac{c_j^{II}}{c_{j,s}}-1\right)}{\left(\sum_{j=1}^{N}\left(\frac{S_{m,j1}^2}{c_{j,s}}\right)\right)} \tag{42}$$

which defines the matrix elements of Eq (17).

## Appendix B: Derivation of $Det\ \mathbf{M_1} = \mathbf{0}$

Here it is shown that $Det\ \boldsymbol{M}_1 = 0$ when the coordinates of the spinodal, Eq (20), are substituted.

$$Det\ \boldsymbol{M_1} = \begin{vmatrix} 2\sum_{\substack{j=1\\j\neq 1}}^{N}(B_{1j}S_{sp,j1}) & 2B_{12} & \cdots & 2B_{1N} \\ 2B_{12} & 2\sum_{\substack{j=1\\j\neq 2}}^{N}(B_{2j}S_{sp,j2}) & \cdots & 2B_{2N} \\ \vdots & \vdots & \ddots & \vdots \\ 2B_{1N} & 2B_{2N} & \cdots & 2\sum_{\substack{j=1\\j\neq N}}^{N}(B_{Nj}S_{sp,jN}) \end{vmatrix}$$

$$= \frac{1}{\prod_{i=1}^{N} S_{sp,i1}} \cdot \begin{vmatrix} \left(2\sum_{\substack{j=1\\j\neq 1}}^{N}(B_{1j}S_{sp,j1})\right)S_{sp,11} & 2B_{12}S_{sp,11} & \cdots & 2B_{1N}S_{sp,11} \\ 2B_{12}S_{sp,21} & \left(2\sum_{\substack{j=1\\j\neq 2}}^{N}(B_{2j}S_{sp,j2})\right)S_{sp,21} & \cdots & 2B_{2N}S_{sp,21} \\ \vdots & \vdots & \ddots & \vdots \\ 2B_{1N}S_{sp,N1} & 2B_{2N}S_{sp,N1} & \cdots & \left(2\sum_{\substack{j=1\\j\neq N}}^{N}(B_{Nj}S_{sp,jN})\right)S_{sp,N1} \end{vmatrix} \quad (43)$$

$$= \frac{1}{\prod_{i=1}^{N} S_{sp,i1}} \cdot \begin{vmatrix} -\left(2\sum_{\substack{j=1\\j\neq 1}}^{N}(B_{1j}S_{sp,j1})\right) & 2B_{12}S_{sp,11} & \cdots & 2B_{1N}S_{sp,11} \\ 2B_{12}S_{sp,21} & -\left(2\sum_{\substack{j=1\\j\neq 2}}^{N}(B_{2j}S_{sp,j1})\right) & \cdots & 2B_{2N}S_{sp,21} \\ \vdots & \vdots & \ddots & \vdots \\ 2B_{1N}S_{sp,N1} & 2B_{2N}S_{sp,N1} & \cdots & -\left(2\sum_{\substack{j=1\\j\neq N}}^{N}(B_{Nj}S_{sp,j1})\right) \end{vmatrix}$$

In the first step each $i$-th row is multiplied by $S_{sp,i1}$, and in the next step Eq (21) is applied. Adding the sum of rows 2 to $N$ to the first row leads to

$$Det\ \boldsymbol{M_1} = \frac{1}{\prod_{i=1}^{N} S_{sp,i1}} \cdot \begin{vmatrix} 0 & 0 & \cdots & 0 \\ 2B_{12}S_{sp,21} & -\left(2\sum_{\substack{j=1\\j\neq 2}}^{N}(B_{2j}S_{sp,j1})\right) & \cdots & 2B_{2N}S_{sp,21} \\ \vdots & \vdots & \ddots & \vdots \\ 2B_{1N}S_{sp,N1} & 2B_{2N}S_{sp,N1} & \cdots & -\left(2\sum_{\substack{j=1\\j\neq N}}^{N}(B_{Nj}S_{sp,j1})\right) \end{vmatrix} = 0 \quad (44)$$

as was to be demonstrated.

## Appendix C: Proof for the minor-rule

In Appendix B it was shown that $Det\ \boldsymbol{M_1} = 0$ when the coordinates of the spinodal (cf. Eq (20)) are substituted. Now consider the (Laplace) determinant expansion by minors along the $i$-th row

$$Det\ \boldsymbol{M}_1 = \sum_{j=1}^{N}\left((-1)^{i+j}\boldsymbol{M}_1(i,j)\{Det\ \boldsymbol{M}_1^{(i,j)}\}\right) = 0 \tag{45}$$

Note that

$$\boldsymbol{M}_1(i,j) = \begin{cases} 2\sum_{\substack{j=1\\ j\neq i}}^{N}\left(B_{ij}S_{sp,ji}\right) & for\ i = j \\ 2B_{ij} & for\ i \neq j \end{cases} \tag{46}$$

It follows that

$$\begin{aligned} Det\ \boldsymbol{M}_1 &= \sum_{j=1}^{N}\left((-1)^{i+j}\boldsymbol{M}_1(i,j)\{Det\ \boldsymbol{M}_1^{(i,j)}\}\right) = \boldsymbol{M}_1(i,i)\{Det\ \boldsymbol{M}_1^{(i,i)}\} + \sum_{\substack{j=1\\ j\neq i}}^{N}\left((-1)^{i+j}\boldsymbol{M}_1(i,j)\{Det\ \boldsymbol{M}_1^{(i,j)}\}\right) \\ &= \left(2\sum_{\substack{j=1\\ j\neq i}}^{N}\left(B_{ij}S_{sp,ji}\right)\right)\{Det\ \boldsymbol{M}_1^{(i,i)}\} + 2\sum_{\substack{j=1\\ j\neq i}}^{N}\left((-1)^{i+j}B_{ij}\{Det\ \boldsymbol{M}_1^{(i,j)}\}\right) \\ &= 2\sum_{\substack{j=1\\ j\neq i}}^{N}\left(\left((-1)^{i+j}\{Det\ \boldsymbol{M}_1^{(i,j)}\} + S_{sp,ji}\{Det\ \boldsymbol{M}_1^{(i,i)}\}\right)B_{ij}\right) = 0 \end{aligned} \tag{47}$$

This must be true for all possible values of $B_{ij}$, which can only be achieved if (for fixed row *i*)

$$\{Det\ \boldsymbol{M}_1^{(i,i)}\} = (-1)^{i+j+1}S_{sp,ij}\{Det\ \boldsymbol{M}_1^{(i,j)}\} \tag{48}$$

The argument can be repeated for the determinant expansion by minors along the *j*-th column, which leads (for fixed column *j*) to

$$\{Det\ \boldsymbol{M}_1^{(j,j)}\} = (-1)^{i+j+1}S_{sp,ji}\{Det\ \boldsymbol{M}_1^{(i,j)}\} \tag{49}$$

By converting $\{Det\ \boldsymbol{M}_1^{(i,j)}\} \xrightarrow[j\ fixed]{} \{Det\ \boldsymbol{M}_1^{(j,j)}\} \xrightarrow[j\ fixed]{} \{Det\ \boldsymbol{M}_1^{(N,j)}\} \xrightarrow[i\ fixed]{} \{Det\ \boldsymbol{M}_1^{(N,N)}\}$, the minor rule, Eq (28), can be obtained.

$$Det\ \boldsymbol{M}_1^{(i,j)} = (-1)^{i+j+1}S_{sp,ji}\left\{(-1)^{i+N+1}S_{sp,iN}\left\{(-1)^{i+N+1}S_{sp,iN}\{Det\ \boldsymbol{M}_1^{(N,N)}\}\right\}\right\} = (-1)^{i+j}S_{sp,iN}S_{sp,jN}\{Det\ \boldsymbol{M}_1^{(N,N)}\} \tag{50}$$

as was to be demonstrated.

## Acknowledgement

The authors acknowledge initial discussions on linear programming with Jelle de Snoo and Rogier Brussee.